\documentclass{iopjournal}

\usepackage{multirow}
\usepackage{amsmath}
\usepackage{comment}
\usepackage[numbers,sort&compress]{natbib}

\newcommand{\red}[1]{\color{black}{#1}\color{black}}

\begin{document}

\articletype{Paper} 

\title{Stochasticity of fatigue failure times in sheared glasses}

\author{
Swarnendu Maity$^{1\dag}$\orcid{0009-0003-8752-6266}, Pushkar Khandare$^{1\dag}$\orcid{0009-0000-5904-2318}, 
Himangsu Bhaumik$^1$\orcid{0000-0003-3893-0775}, Peter Sollich$^2$\orcid{0000-0003-0169-7893},and Srikanth Sastry$^{*,1}$\orcid{0000-0001-7399-1835}
}

\affil{$^1$Jawaharlal Nehru Center for Advanced Scientific Research, Jakkur Campus, Bengaluru 560064, India.}

\affil{$^2$Institute for Theoretical Physics, University of G\"ottingen, Friedrich-Hund-Platz 1, 37077 G\"ottingen, Germany}

\affil{$^\dag$These authors contributed equally to this work.}

\affil{$^*$Author to whom any correspondence should be addressed.}

\email{sastry@jncasr.ac.in}


\begin{abstract} Fatigue failure occurs when a solid is subjected to repeated, cyclic loading. Glasses subjected to cyclic shear deformation have recently been investigated using computer simulations and theoretical models, to characterize and rationalize the dependence of the number of cycles to failure, depending on the properties of the glasses, and the deformation amplitude. The average number of cycles to failure has been observed to diverge as the strain amplitude approaches the so-called fatigue limit from above. In this work, rather than the average times themselves, we investigate by computer simulations the distribution of fatigue failure times, in model glasses subjected to cyclic shear deformation and in an elasto-plastic model.  In particular, we observe in atomistic simulations that the standard deviation of the logarithm of failure times are proportional to their mean values, with the proportionality constant decreasing as the system size increases, indicating a sharper distribution of failure times. Using a finite-element-based elasto-plastic model, we observe similar behavior and perform a system-size analysis showing that the ratio of the standard deviation to the mean tends toward zero in the thermodynamic limit. Such distributions, rather than arising solely from the distribution of disorder in the samples that have been subjected to cyclic deformation, appear to arise from the intrinsic stochasticity of the failure process, which we analyze through a stochastic damage accumulation model. 
\end{abstract}


\section{Introduction}
Understanding the mechanical response of a solid subjected to externally applied stress or deformation has its obvious importance for practical purposes. Typically, solids exhibit elastic behaviour when the applied stress is small. For large enough applied stress or deformation, they exhibit plastic response and eventually fail. Apart from its evident practical importance, the phenomena associated with such failure also present challenges for theoretical understanding, as dynamical transitions in driven systems. Statistical mechanical approaches to such transitions involving failure and fluidization have received considerable interest, particularly in the case of amorphous solids \cite{SethnaAnnuRevMaterRes2017,BonnRevModPhys2017, NicolasRevModPhys2018,rosso_bookchapter_spinglass,Berthier2025,bonfanti2025recentadv}, with analogies to related phenomena in other disordered systems. Among various failure modes (\textit{e.g.} yielding, creep, buckling) that a solid may experience, a particularly interesting and important one is the phenomenon of fatigue failure, wherein a solid, subjected to repeated cycles of stress, or deformation, fails after several cycles of stress or deformation \cite{SureshCUP1998,Christensen2014}. The number of cycles it takes to fail (or, the failure time) depends on the magnitude of the stress or deformation applied in a cycle, and increases upon reduction of magnitude, towards a threshold value known as the fatigue limit. Below the fatigue limit, a solid can sustain an indefinite number of cycles of loading without failing. In addition to the average failure times, the behavior of the distributions of failure times are interesting to investigate, and should be expected to contain valuable information into the mechanisms of failure. In the present work, we consider such distributions, obtained from computer simulations of model glasses subjected to cyclic shear deformation, and an elasto-plastic model that is designed to capture essential features of such sheared glasses.  





The theoretical and computational investigation of yielding behaviour under cyclic shear deformation has drawn considerable attention recently \cite{FioccoPRE2013,Regev2013,Priezjev2013,FioccoIOP2015,LeishanthemNatCom2017,kawasakiPRE16,ParmarPRX2019,BhaumikPNAS2021,YehPRL2020,SastryPRL2021,MunganPRL2021,LiuJCP2022,ParleyPRL2022,CochranPRL2024,PRIEZJEV2021yielding}. Glasses subjected to cyclic shear strain undergo a yielding transition when the amplitude amplitude of strain reaches a critical value. Such a transition is marked by a change in particle dynamics (cycle to cycle) from non-diffusive to diffusive behaviour \cite{FioccoPRE2013,Regev2013,FioccoIOP2015,LeishanthemNatCom2017,kawasakiPRE16,BhaumikJCP2022}, and is  accompanied by strain localization and shear banding above the transition \cite{ParmarPRX2019,BhaumikPRL2022}. 

The nature of the transition also displays a striking dependence on the degree of annealing
of the glasses. Poorly annealed glasses display mechanically induced annealing, reaching lower energies with increasing strain amplitude, till the yield point is reached,  beyond which the energies increase with strain amplitude \cite{LeishanthemNatCom2017}. Well annealed glasses display little change in properties till the yield point is reached, at which discontinuous yielding (with the degree of discontinuity increasing with annealing) takes place. Such qualitative change in behaviour was observed, in computer simulations, to occur for glasses that are obtained from instantaneous quenches from the liquid at equilibrium close to the mode coupling temperature \cite{BhaumikPNAS2021}.  

Fatigue failure under cyclic loading, observed when the applied strain amplitude exceeds the yield amplitude, has been studied extensively in recent work \cite{Bhowmik_2022,BhowmikPRE2022,MaityNatPhys2026}.
The results in \cite{MaityNatPhys2026} show that the average number of cycles required for failure increases rapidly and appears to diverge with power-law exponent $-2$ (in three dimensions) as the strain amplitude approaches the yield point from above, and the exponent value is independent of the degree of annealing of the initial glass. Further, fatigue failure times bear striking relationships with the progressive accumulation of irreversible plastic activity, quantified through measures such as non-affine particle displacements or dissipated energy per cycle. The observed relations between these accumulated plasticity measures and the failure times remain to be understood theoretically. Nevertheless, these results suggest that fatigue failure can be understood as a gradual damage-accumulation process driven by repeated plastic rearrangements. 

In general terms, such cumulative-damage dynamics can be well described using stochastic degradation models \cite{ParkIEEE05,stoch_degradation_review}
in which microscopic irreversible events progressively degrade the system until a failure threshold is reached. In this framework, the accumulated damage is often modeled as a geometric Brownian motion \cite{ParkLDA05} which provides a simple description of monotonic damage accumulation with cycle-to-cycle variability in the degradation rate, and they predict a broad distribution of failure times even under identical loading conditions. Such a broad distribution of the failure time is intrinsically related to the stochastic nature of the failure mechanism in the disordered solids. Different views have been expressed in earlier works regarding the stochastic nature of the location of plastic rearrangements and eventual shear banding, as to whether such locations may be imprinted in the structure inhomogeneities in the initial undeformed glass \cite{FanFalkPRM22,OzawaPRR22,RichardPRMat20,VaibhavPRM23} or are determined dynamically and thus depend on the details of the shear protocol \cite{GendelmanEPL15,Shrivastav2016}. We address this question briefly, and somewhat indirectly in the context of the stochasticity of failure times.

While recent numerical studies on fatigue failure of glasses have primarily focused on the scaling of the mean failure time \cite{KurotaniCommunMat2022,Bhowmik_2022,PRIEZJEV2023Fatigue,MaityNatPhys2026}, the full distribution of fatigue failure times has received comparatively less attention. However, experiments on fatigue in a variety of materials consistently show that failure times are inherently stochastic, even under controlled loading conditions \cite{sobczyk1992,Schijve2009,Raman09fatiguedist,ParkIEEE05,ParkLDA05,elkhoukhi2022,Tridello2023}. Characterizing and rationalizing the failure time distributions is therefore important for developing a statistical mechanical understanding of fatigue failure and for predicting survival lifetimes of materials subjected to cyclic deformation.

In this article, we present results for the distribution of fatigue failure times from simulations of model glasses and an elasto-plastic model. In Sec. \ref{Sec: models}, we first describe the models and methods used. In Sec. \ref{Sec: dist_tf}, we discuss the possible functional forms that may describe well the distribution of failure times and show that the lognormal and the inverse Gaussian distributions  describe the data well. We also show a scaling collapse of distributions of the logarithm of failure times for different strain amplitudes, which illustrates that the widths of these distributions are proportional to the mean value. However, such proportionality does not hold when the system sizes are varies. The dependence of the failure time distributions on system sizes and degrees of annealing is shown in Sec. \ref{Sec: syssize_annealing}. Results from different dynamical evolutions, using iso-configurational runs and a stochastic thermostat, are discussed in Sec. \ref{Sec: different_runs}. Results from elasto-plastic model simulations are shown in Sec. \ref{Sec: results_from_EPM}. In Sec. \ref{Sec: multiplicativedamage} we  describe  to what extent a model of multiplicative accumulation of damage describes our results, 
for both the atomistic and elasto-plastic models. Finally, we summarize our results in Sec. \ref{Sec: conclusions}.




\section{Models and methods}
\label{Sec: models}

\subsection{Atomistic Model Glasses}
We simulate two three-dimensional (3D) model glasses: (a) the 80:20 Kob-Anderson binary mixture of particles with Lennard-Jones (KA-BMLJ) interactions   \cite{kabmlj_ref,LeishanthemNatCom2017} and (b) the  Coslovich-Pastore model (CP), a binary mixture of particles with short-ranged interactions, with network-forming ability, mimicking the properties of silica \cite{CoslovichJPCM2009}. It is computationally challenging to generate results of sufficient quality for the distributions of failure times, as opposed to their mean values. We therefore choose to work with two model glasses with short-ranged interactions. 
Owing to the short-ranged interaction, the computational time is tractable, and this allows us to simulate enough samples to obtain reasonably good statistics of the failure times.

The interaction potential for the KA-BMLJ model is:
\begin{equation}
\begin{aligned}
    U_{\alpha \beta}(r) =& 4\epsilon_{\alpha\beta}\left[ \left(\frac{\sigma_{\alpha\beta}}{r}\right)^{12} - \left(\frac{\sigma_{\alpha\beta}}{r}\right)^{6}
    + c_{0,\alpha\beta} + c_{2,\alpha\beta}\left(\frac{r}{\sigma_{\alpha\beta}}\right)^2 \right], &r<r_{c,\alpha\beta}\\
    =& 0, &r>r_{c,\alpha\beta},
\end{aligned}
\end{equation} and the interaction potential for the CP model is:
\begin{equation}
\begin{aligned}
    U_{\alpha \beta}(r) =& 4\epsilon_{\alpha\beta}\left[ \left(\frac{\sigma_{\alpha\beta}}{r}\right)^{12} -  (1-\delta_{\alpha \beta}) \left(\frac{\sigma_{\alpha\beta}}{r}\right)^{6}
    + c_{0,\alpha\beta} + c_{2,\alpha\beta}\left(\frac{r}{\sigma_{\alpha\beta}}\right)^2 \right], &r<r_{c,\alpha\beta}\\
    =& 0, &r>r_{c,\alpha\beta},
\end{aligned}
\end{equation}
where $\alpha,\beta \in A,B$ denotes the particle type. For the CP model, the types $A$ and $B$ represent Silicon and Oxygen, respectively, and $\delta_{\alpha \beta}$ represents the Kronecker delta. 

The parameters for the KA-BMLJ potential are $\epsilon_{AB}/\epsilon_{AA} = 1.5$, $\epsilon_{BB}/\epsilon_{AA} = 0.5$, $\sigma_{AB}/\sigma_{AA} = 0.8$, $\sigma_{BB}/\sigma_{AA} = 0.88$, $r_{c,\alpha\beta} = 2.5\sigma_{\alpha\beta}$, while for the CP model $\epsilon_{AB}/\epsilon_{AA} = 24.0$, $\epsilon_{BB}/\epsilon_{AA} = 1.0$, $\sigma_{AB}/\sigma_{AA} = 0.49$, $\sigma_{BB}/\sigma_{AA} = 0.85$, $r_{c,\alpha\beta} = 2.5\sigma_{\alpha\beta}$. The terms $c_{0\alpha\beta}$ and $c_{2\alpha\beta}$ are chosen so that both the potential and the force go to zero at $r_{c,\alpha\beta}$. Units of length, energy, and time are $\sigma_{AA}, \epsilon_{AA}$, and $\sqrt{{\sigma_{AA}^2}/{\epsilon_{AA}}}$ respectively. We use an integration time step of $0.005$ for equilibration.
For both the KA-BMLJ and the CP model, initial glasses at different degrees of annealing are prepared \textit{via} molecular dynamics simulations at constant volume and temperature (with the Nos\'e-Hoover thermostat). Configurations are sampled from equilibrated trajectories corresponding to parent temperatures $T_p \in [0.435,1.0]$ for KA-BMLJ, at a reduced density $1.2$, and $T_p = 0.36$ for the CP model, at a reduced density $1.93$. These configurations are subjected to energy minimization, resulting in so-called inherent structures that are used as initial glass samples. For the KA-BMLJ model, initial glass samples with ultra-low degree of annealing ($e_{IS} < -7.00$) are prepared by mechanical annealing, as described in Ref. \cite{DasJCP22}. \footnote{However, we note that comparable low temperature configurations were also obtained in \cite{DasJCP22} through conventional MD simulations.} We implement strain-controlled cyclic shear at finite temperatures and strain rates \textit{via} the Nos\'e-Hoover thermostatted SLLOD \cite{EvansPRA1984} equations of motion. Further details on the cyclic shear protocol can be found in \cite{MaityNatPhys2026}. For the KA-BMLJ model, we consider $64$ samples for the system size $N=4000$, and $32$ samples for $N=16000, 64000, 128000$, for different strain amplitudes and annealing levels. In addition, we consider $400$ samples for $N=4000$ for a particular strain amplitude $\gamma_{max}=0.086$ to obtain better statistics. For the CP model, we consider $16-32$ samples. All numerical simulations, including molecular dynamics, cyclic shear, and energy minimization using the conjugate gradient algorithm, are performed using modified (in-house)  version of LAMMPS \cite{LAMMPS}. 

Application of repeated cycles of strain drives the system to a steady state which can be characterized by the cycle-to-cycle displacement of particles, which either vanishes (for small strain amplitude $\gamma_{max}$) or reaches finite values (for amplitudes $\gamma_{max}>\gamma_{max}^Y$, $\gamma_{max}^Y$ being the yield strain amplitude). The system exhibits a transition from an {\it absorbing} to a {\it diffusive} state at the yield strain amplitude \cite{FioccoPRE2013, LeishanthemNatCom2017,MaityNatPhys2026}. In this work we focus on the regime $\gamma_{max}>\gamma_{max}^Y$, where glasses subject to cyclic deformation transition to a diffusive steady state after a finite number of strain cycles, which we refer to as the failure time $t_f$. To identify the failure time, we monitor the per-particle potential energy $U/N$, which changes sharply when the system fails \cite{ParmarPRX2019,MaityNatPhys2026}. The failure time $t_f$ is estimated to be the midpoint of the transformation, using a fit to a sigmoid function as described in Ref. \cite{MaityNatPhys2026}.

\subsection{Elastoplastic Model (EPM)}

EPMs (see \cite{NicolasRevModPhys2018} for a comprehensive review) are coarse-grained mesoscopic models that have been successful in capturing the phenomenology of driven amorphous solids. These models, owing to their coarse-grained nature, remain computationally accessible for large system sizes and large number of samples. This fact has been exploited recently \cite{pollard_22,rossi_22} to examine the nature of the ductile-to-brittle transition via extensive finite-size simulations. We use a recently developed energy-landscape based elasto-plastic model \cite{khandare2026} which is a spatially resolved extension of a single-site \textit{mesostate} model described in \cite{SastryPRL2021}. The single-site model in \cite{SastryPRL2021} was motivated by the observation of mechanically induced annealing when glasses are cyclically sheared \cite{LeishanthemNatCom2017,ParmarPRX2019,BhaumikPNAS2021}. In this model, wherein the amorphous solid is described as a regular array of mesoscopic blocks (or sites), the elastic interactions between the constituent sites are calculated \textit{via} the finite element method. Mesoscopic blocks (or mesostates) are arranged in a regular square grid of size $L\times L$ in two dimensions, and strain-controlled periodic boundary conditions are assumed. We consider three system sizes: $L=128,256,512$. The mesostates respond elastically when deformed; they follow a quadratic form for the local elastic energy, $E_i = E_{0i} + \frac{\mu_i}{2}(\gamma_i - \gamma_{0i})^2$, where the subscript $i$ refers to the $i$th site in the grid, and we set $\mu_i =2$ for all sites. The local \textit{plastic} strain ($\gamma_0$) denotes the stress-free strain of the mesostate. The local stress-free energy is termed as $E_0$ and is chosen to lie within $(-1,0)$. It is assumed that any given mesostate is stable when the local strain in that plaquette lies within $\gamma_0 \pm \sqrt{-E_0}$, that is, deeper (shallower) minima are stable over a larger (smaller) range of strain. Once a block becomes unstable the new value for $E_0$ is sampled from the density of states which is assumed to follow a Gaussian distribution with mean $-0.5$ and standard deviation $0.1$. The new value of $\gamma_0$ is taken to be $\gamma_0 \pm (\sqrt{-E_0^{old}} + \sqrt{-E_0^{new}})$, the sign depends on whether the block became unstable on crossing the right (plus) or the left (minus) stability limit. This choice of plastic strain increment implies that the parabolae are stacked edge to edge in either direction. The coupling of the stability range and the energy minimum allows for the intended annealing effect, since mesostates that are too shallow with respect to the cyclic driving amplitude will get replaced by deeper states with larger stability limits. Most prior EPM studies (with some exceptions, e.g., \cite{sollich97,sollich98,Jagla_2007,agoritsas2015,Khirallah_2021,LiuJCP2022}) lack this energy-landscape based approach in that there is no explicit treatment of the energy of a constituent site and its relation to the threshold stress at that site. A novel direction followed in \cite{SastryPRL2021,khandare2026} is that the mesostate energy minimum is equivalent to inherent structure energy in glasses, the distribution of which is known from prior atomistic simulations to be Gaussian \cite{heuer_00,sastry_01}.

Cyclic shear is implemented \textit{via} the athermal quasistatic protocol where the globally imposed strain is incremented to the value where exactly one site becomes unstable. A mesostate transition is implemented and the stress redistribution is calculated via the finite element method. Unstable sites, if any, are made to undergo transitions, till the system reaches a stable state. This process is repeated such that the globally imposed strain traces the path $0\rightarrow \gamma_{max}\rightarrow 0 \rightarrow -\gamma_{max} \rightarrow 0$; this defines a cycle. We consider two annealing levels: a poorly annealed sample corresponding to a parent temperature $T_p = 3.16$ and a well annealed sample corresponding to a parent temperature $T_p = 0.06$. The sample preparation protocol and further details may be found in \cite{khandare2026}. 

This model reproduces the cyclic shear yielding transition; poorly annealed samples get mechanically annealed and reach absorbing states at low driving amplitudes, and they yield beyond a critical amplitude; well annealed samples show negligible mechanical annealing, and they yield at a $T_p$-dependent critical amplitude, following the universal energy curve post-yield. Further, this model shows critical behaviour around the yield amplitude with power-law divergence in time taken to reach the final state on approaching the transition from either side, the failure time exponent being $-1$ for poorly-annealed samples and $-2.2$ for well-annealed samples \cite{khandare2026}. In this model failure is accompanied by the formation of a shear band, with most of the subsequent plastic activity being confined to this region. Thus, an order parameter that can detect a band-like feature, like the one described in \cite{YehPRL2020}, is used to detect the failure time. In practice, the cycle-to-cycle change in plastic strain field is used to construct this order parameter, which is denoted by $H$. We have verified that a threshold of $H > 0.2$ faithfully captures the failure event.

\section{Results}

\subsection{Distribution of failure times}
\label{Sec: dist_tf}

\begin{figure*}[ht!]
        \centering
         \includegraphics[width=.75\linewidth]{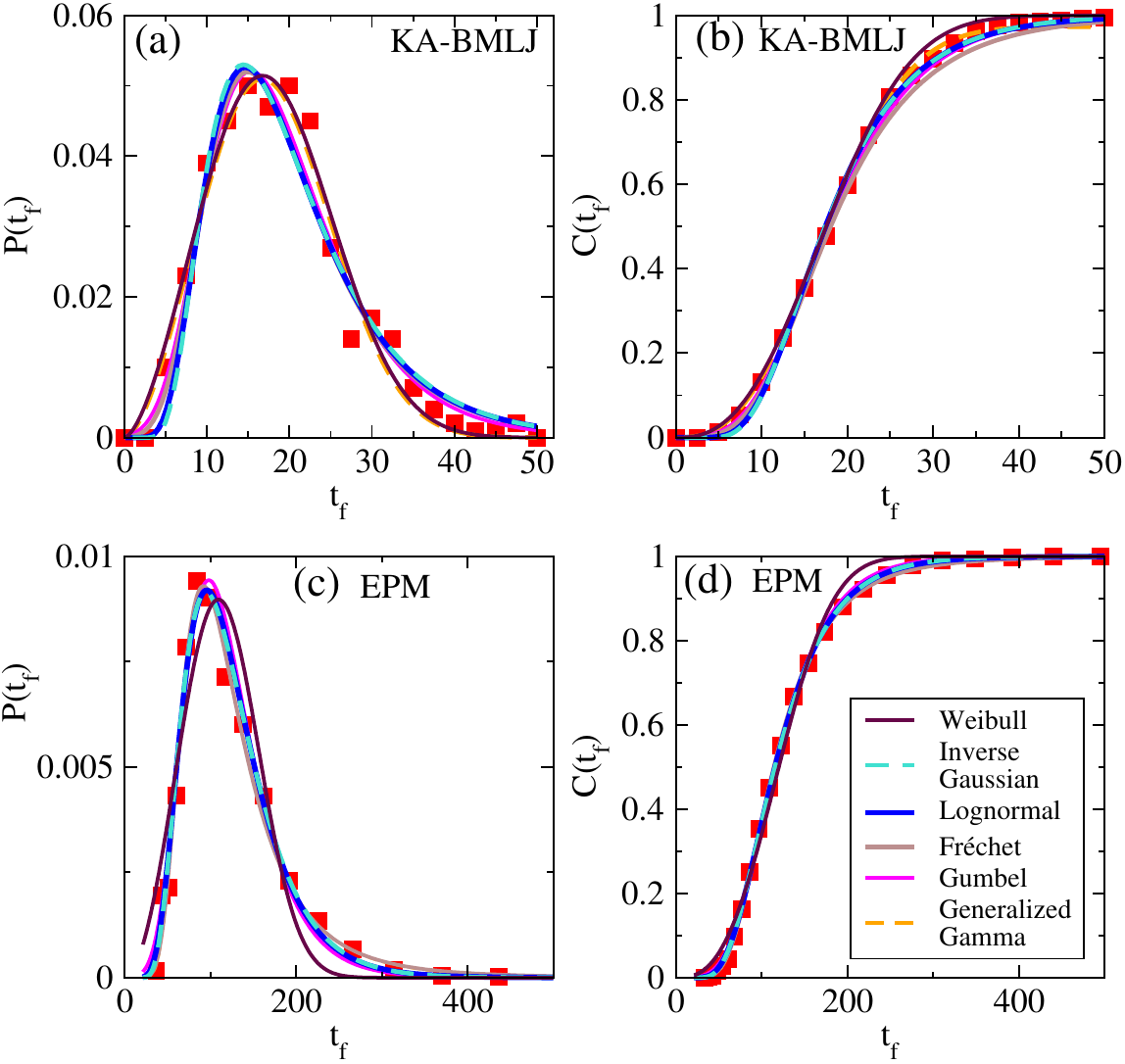}
        \caption{\textbf{Distribution of failure times:} (a) Probability distribution $P(t_f)$ and (b) cumulative distribution $C(t_f)$ of failure times $t_f$ from atomistic simulations where $N=4000$, $e_{IS}=-7.00$ and $\gamma_{max}=0.086$ ($\gamma_{max}^{Y}=0.0718$) using $400$ samples. (c) $P(t_f)$ and (d) $C(t_f)$ from elasto-plastic model (EPM) simulations for a poorly-annealed sample with $L=128$ and $\gamma_{max} = 0.458$ ($\gamma_{max}^{Y}=0.441$) using $1000$ samples. Lines are fits to different functional forms: Weibull, generalized gamma, Fr\'echet, Gumbel, lognormal, and inverse Gaussian [See table \ref{Tab:different_fitting_forms} for the details of each distribution's form and fitting parameters]. All panels share the same color code and the shared legend is presented in (d).}
    \label{fig:failure_time_differnt_fit}
\end{figure*}

When different samples obtained from the same parent temperature ($T_p$) are subjected to the same strain amplitude $\gamma_{max}$, they exhibit a wide variability in their failure lifetimes. Fig. \ref{fig:failure_time_differnt_fit}(a), shows the probability distribution of failure times, $P(t_f)$, obtained from $400$ independent samples of system size $N=4000$, prepared at $T_p=0.435$ (corresponding to $e_{IS}=-7.00$), and subjected to cyclic shear of maximum strain amplitude of $\gamma_{max}=0.086$ ($\gamma_{max}^{Y}=0.0718$). The corresponding cumulative distribution $C(t_f)$ is shown in Fig. \ref{fig:failure_time_differnt_fit}(b). Analogous results from elasto-plastic model simulations are presented in Fig. \ref{fig:failure_time_differnt_fit}(c) and Fig. \ref{fig:failure_time_differnt_fit}(d) for a poorly annealed sample with $L=128$, sheared at $\gamma_{max} = 0.458$ ($\gamma_{max}^{Y}=0.441$). The data is obtained from $1000$ samples. Further results from the elasto-plastic model simulations are presented in Section \ref{Sec: results_from_EPM}.

In the fatigue failure literature, different distribution functions have been considered as descriptions of the observed failure time distributions \cite{Schijve2009,McPherson2019}. The most popular choices are Weibull \cite{SureshCUP1998,weibull_book} and Lognormal distributions \cite{Limpert2001,Brot2019,BhowmikPRE2022_creep}. Widely used, physically motivated distributions which can be derived from the first passage time of an underlying degradation model, are the fatigue life distribution (also known as the Birnbaum-Saunders distribution \cite{Birnbaum_Saunders_1969,kundu_bs_2019}) and the inverse gaussian distribution \cite{ParkIEEE05,ParkLDA05}. It has been shown that the fatigue life distribution is an approximation of the inverse Gaussian distribution \cite{bhattacharyya1982}, and consequently we will only consider the inverse Gaussian distribution in our discussion. We also consider other, less physically motivated, distributions that are used to fit lifetime data, the Fr\'echet \cite{frechet} and the Gumbel distributions \cite{Andersonijfatigue2018} (like the Weibull distribution these are extreme value distributions), and the generalized Gamma distribution \cite{stacy1962,Lawless1980}, which includes many known distributions (Weibull, exponential, Gamma, etc.) as special cases for specific parameter choices. We compare these different forms for our simulation results in Table \ref{Tab:different_fitting_forms}. We observe that all the forms fit the data comparably well. We note, for reference in the subsequent discussions, that the lognormal and inverse Gaussian distributions overlap strongly and are essentially indistinguishable.

In the following, we use cumulative distribution functions for better clarity. We will also concentrate on the lognormal and the inverse Gaussian forms for reasons we explain later.

\begin{table}

\centering
\begin{tabular}{l c r }
\hline
Distribution & Optimal parameters & RRMSE\\
\hline
Weibull:\\
$P(t_f)=\frac{k}{\lambda}\left(\frac{t_f}{\lambda}\right)^{k-1}\exp\left[-\left(\frac{t_f}{\lambda}\right)^k\right]$ & \red{\multirow{2}{*}{\shortstack{MD: $k=2.59$, $\lambda=20.2$\\EPM: $k=2.89$, $\lambda=126.84$}}} & \red{\multirow{2}{*}{\shortstack{MD: $0.027$\\EPM: $0.055$}}} \\\\
$C(t_f)=1-\exp\left[-\left(\frac{t_f}{\lambda}\right)^k\right]$ & \red{\multirow{2}{*}{\shortstack{MD: $k=2.51$, $\lambda=20.9$\\EPM: $k=2.78$, $\lambda=137.09$}}} & \red{\multirow{2}{*}{\shortstack{MD: $0.0034$\\EPM: $0.0089$}}} \\\\
\multicolumn{3}{l}{Mean: $\lambda\Gamma(1+1/k)$, Variance: $\lambda^2\left[\Gamma\left(1+2/k\right)-\left(\Gamma\left(1+1/k\right)\right)^2\right]$}\\
\hline
Inverse Gaussian:\\
$P(t_f)=\sqrt{\frac{\lambda}{2\pi t_f^3}}\exp{\left[-\frac{\lambda (t_f-\mu)^2}{2\mu^2t_f}\right]}$ & \red{\multirow{2}{*}{\shortstack{MD: $\mu=19.19$, $\sigma=94.32$\\EPM: $\mu=124.79$, $\sigma=676.35$}}} & \red{\multirow{2}{*}{\shortstack{MD: $0.038$\\EPM: $0.023$}}} \\\\
\multirow{2}{*}{\shortstack[l]{$C(t_f)=\frac{1}{2}\left[1+\text{erf}\left(\sqrt{\frac{\lambda}{2 t_f}}\left(\frac{t_f}{\mu}-1\right)\right)\right]$ \\$+ \exp{\left(\frac{2\lambda}{\mu}\right)}\frac{1}{2}\left[1+\text{erf}\left(-\sqrt{\frac{\lambda}{2 t_f}}\left(\frac{t_f}{\mu}+1\right)\right)\right]$}} & \red{\multirow{2}{*}{\shortstack{MD: $\mu=20.26$, $\lambda=87.76$\\EPM: $\mu=126.85$, $\lambda=650.38$}}} & \red{\multirow{2}{*}{\shortstack{MD: $0.0047$\\EPM: $0.0020$}}} \\\\\\\\
\multicolumn{3}{l}{Mean: $\mu$, Variance: $\mu^3/\lambda$}\\
\hline
Lognormal:\\
$P(t_f)=\frac{1}{t_f \sigma\sqrt{2\pi}}e^{-\frac{\left(\ln{t_f} - \mu \right)^2}{2\sigma^2}}$ & \red{\multirow{2}{*}{\shortstack{MD: $\mu=2.9$, $\sigma=0.46$\\EPM: $\mu=4.74$, $\sigma=0.41$}}} & \red{\multirow{2}{*}{\shortstack{MD: $0.036$\\EPM: $0.024$}}} \\\\
$C(t_f)=\frac{1}{2}\left[1+\text{erf}\left(\frac{\ln{t_f}-\mu}{\sigma\sqrt{2}}\right)\right]$ & \red{\multirow{2}{*}{\shortstack{MD: $\mu=2.86$, $\sigma=0.44$\\EPM: $\mu=4.75$, $\sigma=0.43$}}} & \red{\multirow{2}{*}{\shortstack{MD: $0.0044$\\EPM: $0.0023$}}} \\\\
\multicolumn{3}{l}{Mean: $\exp\left(\mu+\sigma^2/2\right)$, Variance: $\left[\exp\left(\sigma^2\right)-1\right]\exp\left(2\mu+\sigma^2\right)$}\\
\hline
Fr\'echet:\\
$P(t_f)=\frac{\alpha}{s}\left(\frac{t_f-m}{s}\right)^{-1-\alpha}\exp\left[-\left(\frac{t_f-m}{s}\right)^{-\alpha}\right]$ & \red{\multirow{2}{*}{\shortstack{MD: $\alpha=246290$, $s=1739652$ $m=-1739636$\\EPM: $\alpha=4.68$, $s=189.41$ $m=-89.81$}}} & \red{\multirow{2}{*}{\shortstack{MD: $0.029$\\EPM: $0.020$}}} \\\\
$C(t_f)=\exp\left[-\left(\frac{t_f-m}{s}\right)^{-\alpha}\right]$ & \red{\multirow{2}{*}{\shortstack{MD: $\alpha=960$, $s=6470$ $m=-6455$\\EPM: $\alpha=7.2$, $s=285.96$ $m=-186.19$}}} & \red{\multirow{2}{*}{\shortstack{MD: $0.0027$\\EPM: $0.0017$}}} \\\\
\multicolumn{3}{l}{Mean: $m+s\Gamma(1-1/\alpha)$ for $\alpha>1$, Variance: $s^2\left[\Gamma\left(1-2/\alpha\right)-\left[\Gamma\left(1-1/\alpha\right)\right]^2\right]$ for $\alpha>2$; Otherwise, $\infty$ }\\
\hline
Gumbel:\\
$P(t_f)=\frac{1}{\beta}\exp\left[-\frac{t_f-\mu}{\beta}-\exp\left(-\frac{t_f-\mu}{\beta}\right)\right]$ & \red{\multirow{2}{*}{\shortstack{MD: $\mu=15.31$, $\beta=7.06$\\EPM: $\mu=98.29$, $\beta=40$}}} & \red{\multirow{2}{*}{\shortstack{MD: $0.029$\\EPM: $0.029$}}} \\\\
$C(t_f)=\exp\left[-\exp\left(-\frac{t_f-\mu}{\beta}\right)\right]$ & \red{\multirow{2}{*}{\shortstack{MD: $\mu=15.14$, $\beta=6.74$\\EPM: $\mu=101.34$, $\beta=41.7$}}} & \red{\multirow{2}{*}{\shortstack{MD: $0.0026$\\EPM: $0.0038$}}} \\\\
\multicolumn{3}{l}{Mean: $\mu+\beta \gamma$, Variance: $\pi^2\beta^2/6$}\\
\hline
Generalized gamma:\\
$P(t_f)=\frac{p/a}{\Gamma\left(d/p\right)}\left(\frac{t_f}{a}\right)^{d-1}e^{-\left(\frac{t_f}{a}\right)^p}$ & \red{\multirow{2}{*}{\shortstack{MD: $p=1.64$, $a=12.3$, $d=3.51$\\EPM: $p=-0.92$, $a=817.65$, $d=-5.93$}}} & \red{\multirow{2}{*}{\shortstack{MD: $0.024$\\EPM: $0.020$}}} \\\\
$C(t_f)=\frac{\gamma\left(d/p, \left(x/a\right)^p\right)}{\Gamma\left(d/p\right)}$ & \red{\multirow{2}{*}{\shortstack{MD: $p=1.32$, $a=8.29$, $d=4.07$\\EPM: $p=-0.61$, $a=9699.1$, $d=-9.25$}}} & \red{\multirow{2}{*}{\shortstack{MD: $0.0016$\\EPM: $0.0014$}}} \\\\
\multicolumn{3}{l}{Mean: $a\frac{\Gamma\left(\left(d+1\right)/p\right)}{\Gamma\left(d/p\right)}$, Variance: $a^2\left(\frac{\Gamma\left(\left(d+2\right)/p\right)}{\Gamma\left(d/p\right)} - \left(\frac{\Gamma\left(\left(d+1\right)/p\right)}{\Gamma\left(d/p\right)}\right)^2\right)$}\\
\hline
\end{tabular}

\caption{Fitting parameters and relative root mean square error ($RRMSE = \sqrt{\frac{ \frac{1}{n}\sum_{i=1}^n \left({y_{i}-\bar{y}_{i}}\right)^2}{\sum_{i=1}^n \bar{y}_{i}^2}}$, where $y_i$ are the actual data points and $\bar{y}_{i}$ are fit values) obtained by fitting the probability distribution $P(t_f)$ and cumulative distribution $C(t_f)$ of failure times to different fitting forms as shown in Fig. \ref{fig:failure_time_differnt_fit}. \red{MD represents fitting coefficients and RRMSE values obtained from molecular dynamics simulation data (Fig. 1(a,b)), whereas EPM represents the corresponding data from the elasto-plastic model (Fig. 1(c,d))}. Here $\Gamma(\cdot)$ represents the gamma function, $\gamma(\cdot,\cdot)$ represents the lower incomplete gamma function, and erf$(\cdot)$ represents the error function.
$\gamma$ is Euler's constant.
}
\label{Tab:different_fitting_forms}
\end{table}

\begin{figure*}[ht!]
        \centering
        \includegraphics[width=.75\linewidth]{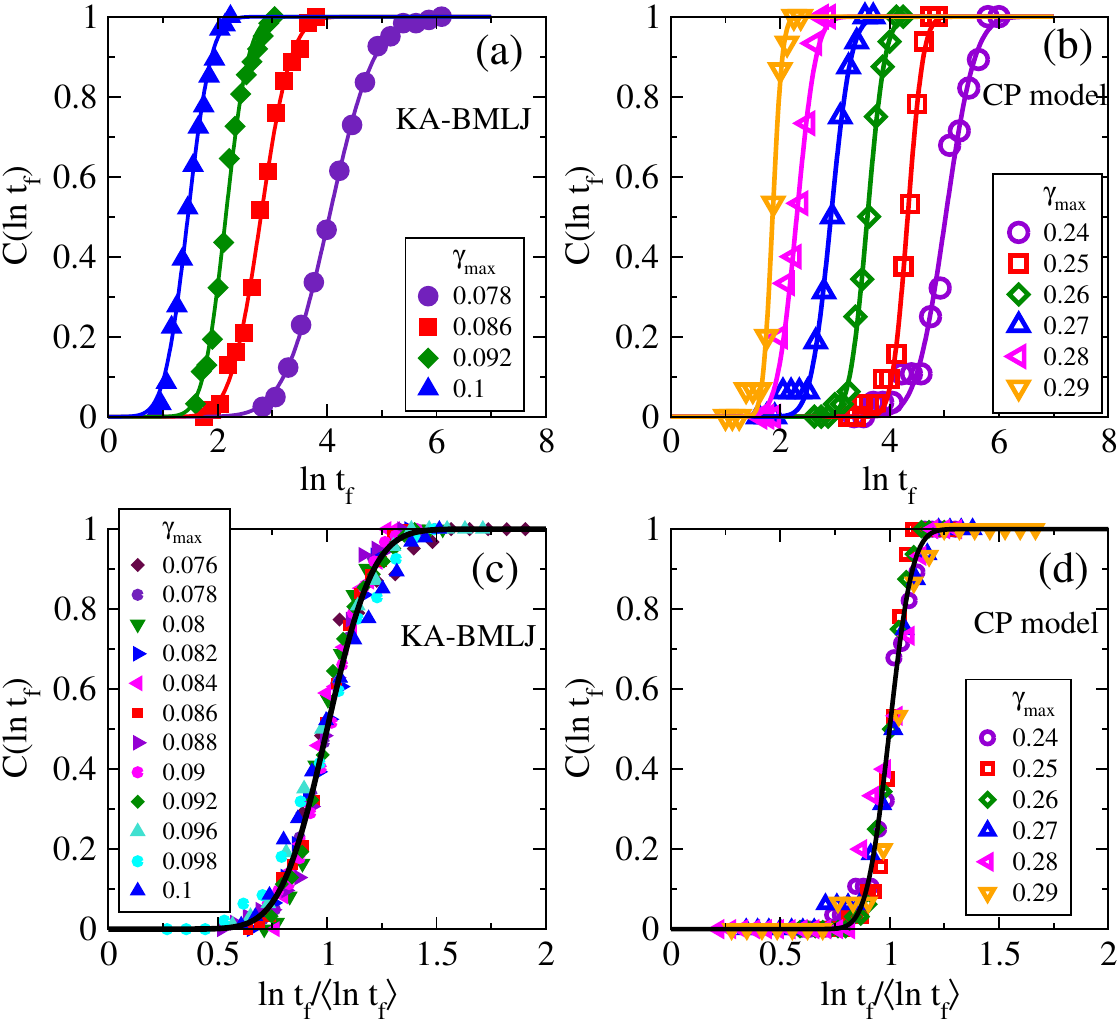}        
    \caption{\textbf{Scaling of the distribution of the failure times:} The cumulative distribution of $\ln{t_f}$ for (a) KA-BMLJ model with $N=4000$, $e_{IS}=-7.00$ $(T_p=0.435)$ and (b) CP model with $N=46656$, $T_p=0.36$ for several strain amplitudes. The solid lines represent fits to the Gaussian distribution. Collapse of data for (c) KA-BMLJ model and (d) CP model for different $\gamma_{max}$ when $\ln{t_f}$ (the $x$-axis) is rescaled with its mean value $\langle\ln{t_f}\rangle$. The solid lines are Gaussian fits to the data points across different $\gamma_{max}$. 
    }
    \label{fig:scaling_distribution_tf}
\end{figure*}


In \cite{BhowmikPRE2022_creep}, failure times were argued to follow a lognormal distribution (or correspondingly, the logarithm of failure time follows a normal distribution) for creep failure under tensile stress. In our simulation of fatigue failure, we also observe that the failure time distribution can be well described by a lognormal distribution for different classes of glasses, \textit{viz.} KA-BMLJ model and CP model of silica, as shown in Fig. \ref{fig:scaling_distribution_tf}(a) and (b) respectively, for different strain amplitudes $\gamma_{max}$. The mean and standard deviation of $\ln{t_f}$ is denoted by $\mu=\langle\ln{t_f}\rangle$ and $\sigma=\sqrt{\langle(\ln{t_f})^2\rangle-\langle\ln{t_f}\rangle^2}$.

Another interesting feature, observed in \cite{BhowmikPRE2022_creep}, is that the distribution across different $\gamma_{max}$ collapses if we scale the $x$-axis $\ln{t_f}$ with the mean value $\mu=\langle\ln{t_f}\rangle$. This data collapse for fatigue failure is shown for the KA-BMLJ model and the CP model in Fig. \ref{fig:scaling_distribution_tf}(c) and (d), respectively. The scaled cumulative distribution, which is shown as the solid lines in \ref{fig:scaling_distribution_tf}(c) and (d), follows:
\begin{equation}\label{eq:scaling_dist_tf}
    C(\ln{t_f}) = \frac{1}{2} \left[1+ \mathrm{erf}{\left(\frac{x - 1}{\sigma'\sqrt{2}}\right)}\right],
\end{equation}
where $x=\ln{t_f} / \langle \ln{t_f} \rangle$, $\sigma' = \sigma / \mu$. 
A constant value of $\sigma'$ across different $\gamma_{max}$, as apparent from the data collapse obtained,  indicates that the mean and standard deviation of $\ln{t_f}$ are proportional. 

\subsection{Dependence of failure time on system size and degree of annealing} 
\label{Sec: syssize_annealing}

\begin{figure*}[ht!]
        \centering
        \includegraphics[width=.98\linewidth]{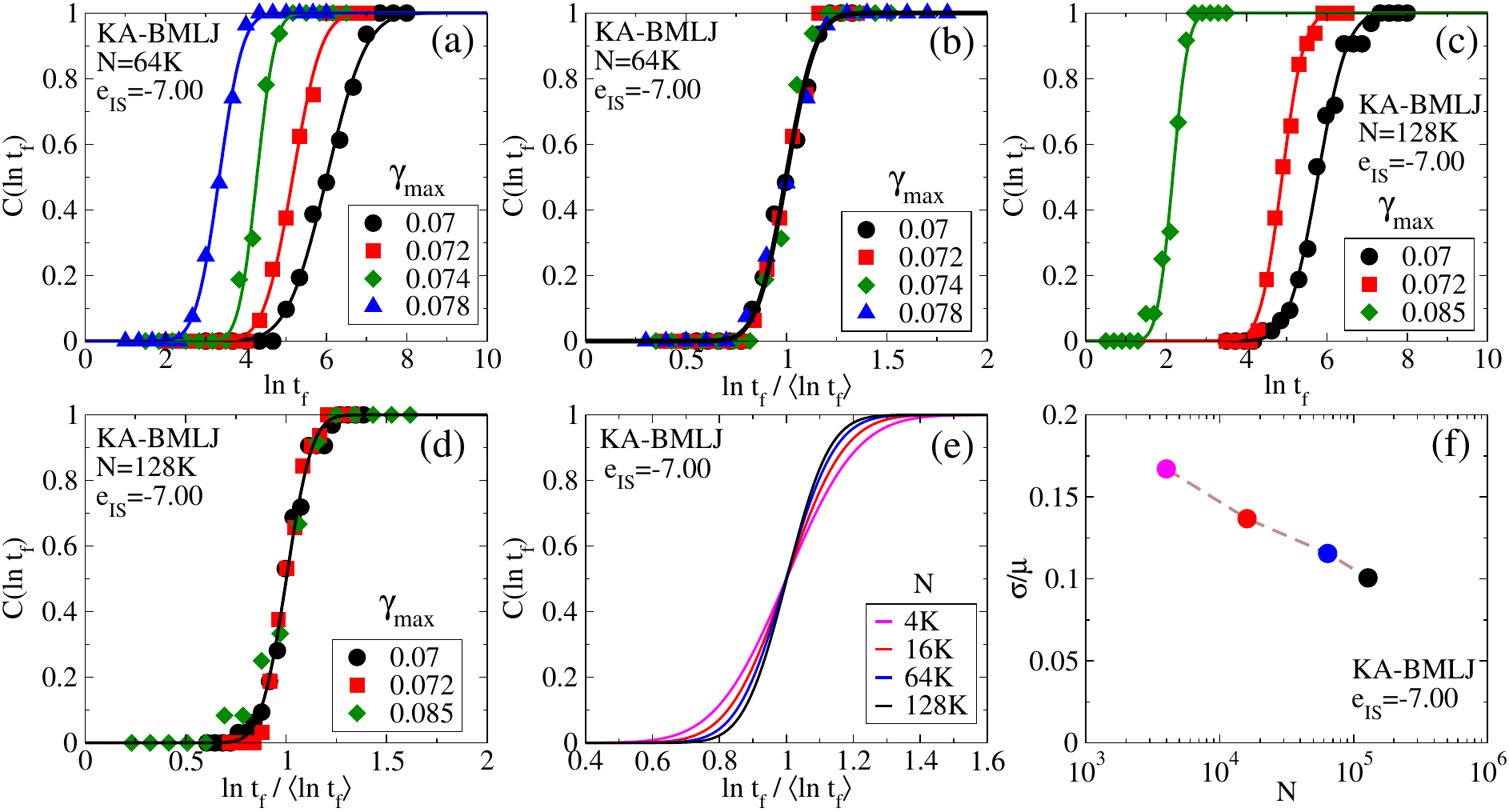}   
    \caption{\textbf{Distribution of failure times for different system sizes (KA-BMLJ):} (a) Cumulative distribution of logarithm of failure times $C(\ln{t_f})$ for a system of size $N=64000$ with $e_{IS}=-7.00$ for different strain amplitudes. Lines are fit to the normal distribution. (b) Collapse of data for different $\gamma_{max}$ when $\ln{t_f}$ is rescaled with its mean value. The solid line through data points represents Eq. (\ref{eq:scaling_dist_tf}).
    (c) and (d) show the same quantities as in (a) and (b), respectively, but for a larger system size of $N=128000$. (e) Scaled cumulative distribution functions (obtained from the fitting) for four different system sizes: $N=4000$, $16000$, $64000$, and $128000$. (Data points are suppressed for the clarity). The distributions become progressively sharper as the system size increases.
    (f) Scaled standard deviation $\sigma_{\ln{t_f}}/\mu_{\ln{t_f}}$ \textit{vs} system size. 
    }
    \label{fig:distribution_tf_systemsize}
\end{figure*}

\begin{figure*}[ht!]
        \centering
        \includegraphics[width=.9\linewidth]{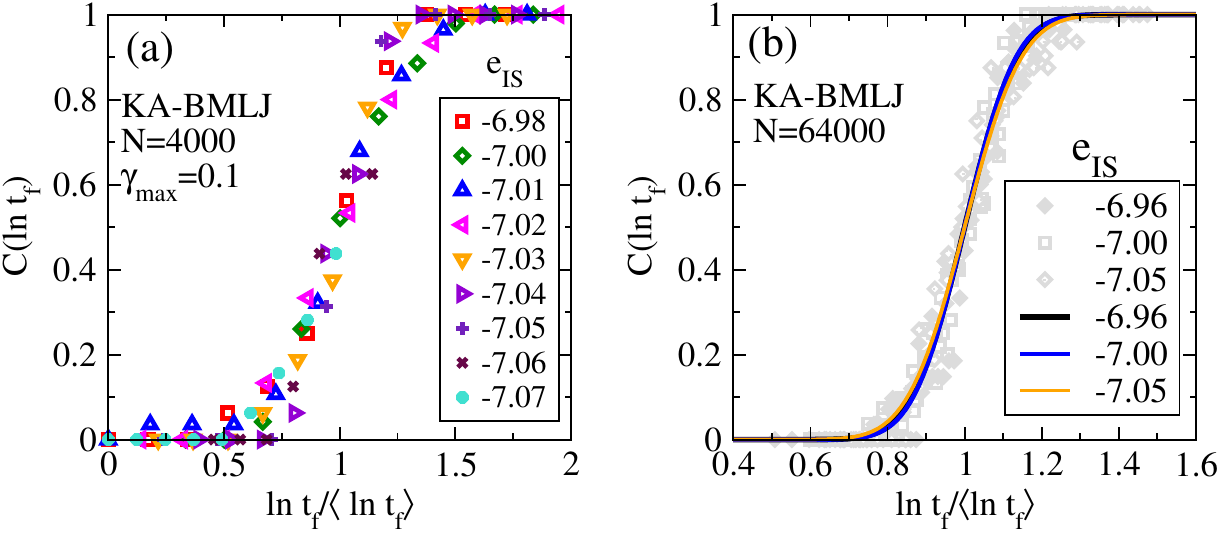}
    \caption{\textbf{Scaling of failure time distributions for different degrees of annealing:} (a) Cumulative distribution as a function of scaled variable $\ln t_f/\langle \ln t_f \rangle $ for $N=4000$, $\gamma_{max}=0.1$, and several degrees of annealing. (b) $C(\ln t_f)$ {\it  vs.} $\ln t_f/\langle \ln t_f \rangle $ for $N=64000$ for three degrees of annealing ($e_{IS}=-6.96, -7.00, -7.05$). Solid lines are fit to the data for the respective degree of annealing. For $N=64000$, data are collected across different $\gamma_{max}$ \red{(for $e_{IS}=-6.96$ and $-7.00$ we consider $\gamma_{max}=0.07, 0.072$, and $0.074$; for $e_{IS}=-7.05$ we consider $\gamma_{max}=0.083, 0.084$, and $0.086$).}
    }
    \label{fig:distribution_tf_anneal}
\end{figure*}

We further investigate the system size dependence of this scaling behaviour. In Fig. \ref{fig:distribution_tf_systemsize}(a-d), we show that the data collapse across different $\gamma_{max}$ holds for different system sizes, for a fixed degree of annealing $e_{IS}=-7.00$ $(T_p=0.435)$. But the standard deviation of the scaled distribution decreases with an increase in system size, as shown in Fig. \ref{fig:distribution_tf_systemsize}(e). Variation of $\sigma' = \sigma/\mu$ for different system sizes is shown in Fig. \ref{fig:distribution_tf_systemsize}(f). 


The cumulative distribution functions across different degrees of annealing for a fixed strain amplitude of $\gamma_{max}=0.1$ and $N=4000$ is shown in Fig. \ref{fig:distribution_tf_anneal}(a). We do not observe a systematic change with the degree of annealing, though the data is not good enough to draw a conclusion from this system size. In Fig. \ref{fig:distribution_tf_anneal}(b), we show the cumulative distribution functions for different degrees of annealing for $N=64000$. 
For each degree of annealing, the scaled distribution is obtained from data collected over several values of $\gamma_{max}$. The solid lines are fits for different degrees of annealing across different $\gamma_{max}$. We do observe that the scaled distributions coincide and do not show dependence on the degree of annealing.

\subsection{Failure times for different dynamics} \label{Sec: different_runs}

\begin{figure*}[ht!]
        \centering
    \includegraphics[width=.98\linewidth]{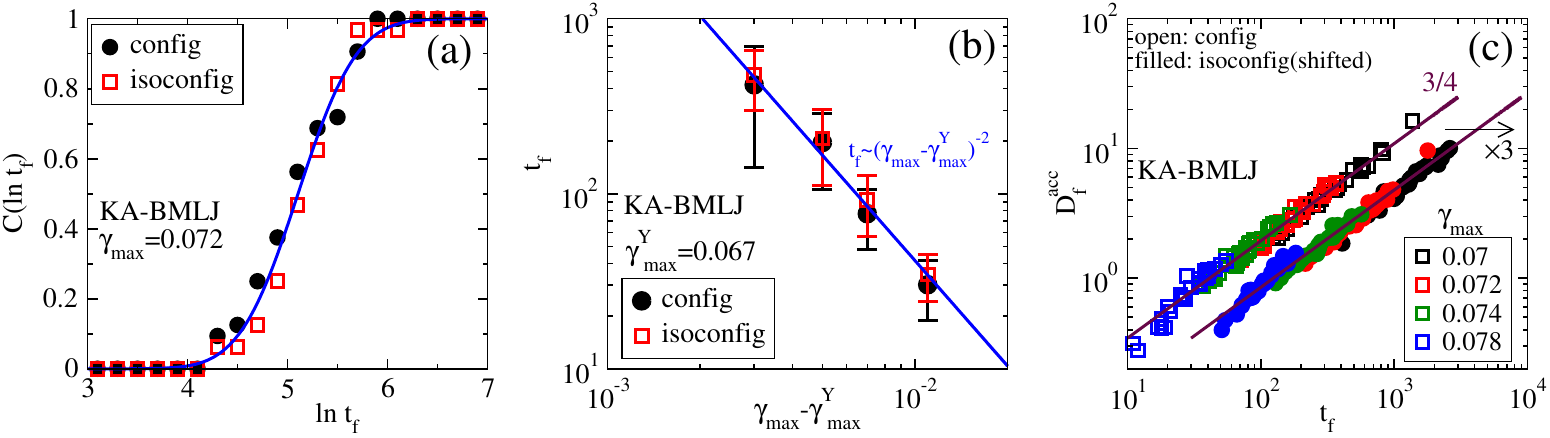}
    \caption{\textbf{Results for different initial conditions:} (a) The cumulative distribution of failure times for runs starting with different initial configurations and with the same initial configurations with different initial velocities. (b) Mean failure time against $\gamma_{max}-\gamma_{max}^Y$. Solid line indicates power law behaviour with exponent $-2$. The error bars represent the sample-to-sample standard deviation of the failure times. (c) Accumulated damage till failure $D^{acc}_f$ grows with failure time $t_f$ with $3/4$ exponent for both cases. For clarity, data for different velocities are shifted towards the right. ($N=64000$, $e_{IS}=-7.00$.) }
    \label{fig:same_config_diff_velocity}
\end{figure*}

So far, we have considered an ensemble of independent configurations obtained from equilibrated liquids at the parent temperature. One possible interpretation of the broad distribution of failure times is that they arise from the {\it structural} stochasticity inherent in such ensembles. Alternately, the stochasticity arises from the dynamical evolution under shear. We test this possibility by considering cyclic shear simulations starting with the same initial configuration, but with independent initializations of the velocities (or, an {\it iso-configurational} ensemble \cite{WidmerCooperPRL2004}).

In Fig. \ref{fig:same_config_diff_velocity}(a), we show the cumulative distribution of failure times for an  ensemble of initial configurations, and an iso-configurational ensemble of velocity initializations, for one strain amplitude. These distributions strikingly display very similar mean and variance. In  Fig. \ref{fig:same_config_diff_velocity}(b), the average failure times, with error bars, are shown for the two cases, for a range of strain amplitudes, which show that these two ensembles of initial conditions lead to very similar average values of the failure time. Finally, in Fig. \ref{fig:same_config_diff_velocity}(c), we show the data collapse of accumulated damage (measured as the dissipated work, the stress-strain loop area, during the cycle of strain) {\textit vs.} failure times \cite{MaityNatPhys2026}. We find the same power-law scaling behaviour $D_{f}^{acc}\sim t_f^{3/4}$ for both the cases. These results strongly support the idea that the relevant stochasticity is, to a significant extent, inherent in the dynamical evolution of the system that contributes to a distribution of failure times.

\subsection{Failure time distributions from simulations of the elasto-plastic model}
\label{Sec: results_from_EPM}

\begin{figure*}[ht!]
    \centering
    \includegraphics[width=.98\linewidth]{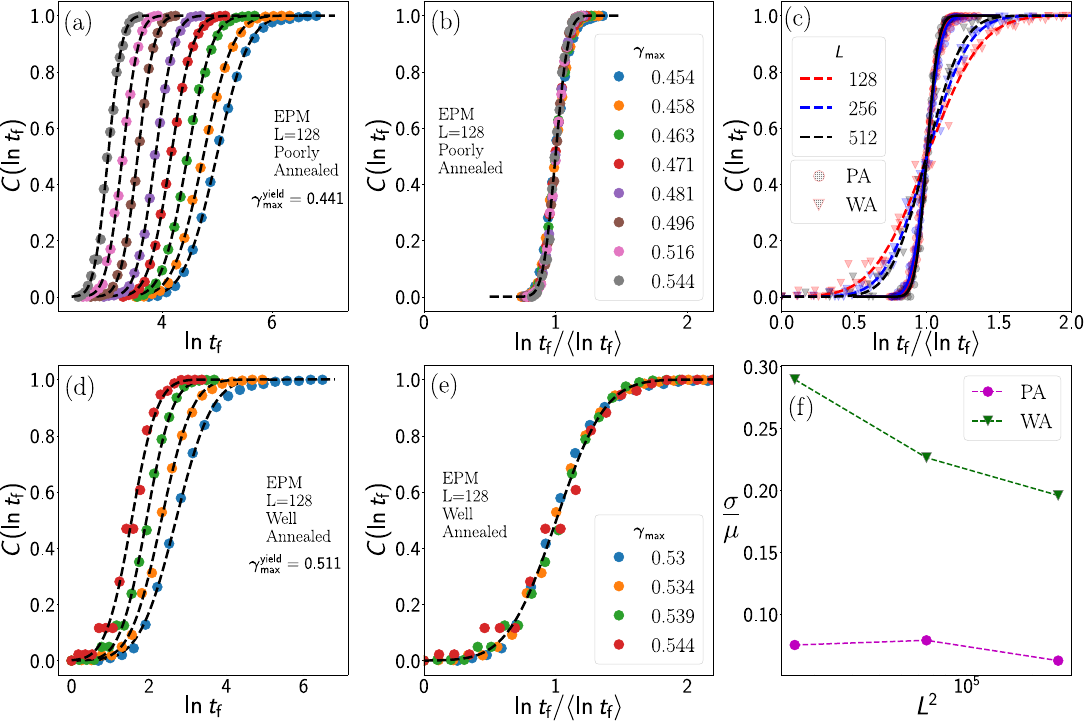}
    \caption{\textbf{Distribution of failure times from EPM:} Cumulative distribution of logarithm of failure times, for system size $L=128$, for different $\gamma_{max}$ is plotted for (a) poorly annealed and (d) well annealed sample. Dashed lines are fit to the normal distribution. (b,\red{e}) Data collapse is observed when the $x$-axis is scaled with the mean value. Panels (a,b) share the same legend (denoted in panel (b)), and similarly the shared legend for panels (d,e) are denoted in panel (e). (c) The scaled distributions are plotted for three system sizes, for both the poorly (circles) and well annealed (inverted triangles) sample. Solid lines are normal fits to the data for poorly annealed sample while dashed lines are normal fits for the well-annealed case. (f) The ratio of standard deviation ($\sigma$) and the mean ($\mu$) extracted from the normal fits are plotted as a function of system size, for both the poorly and well annealed sample. Data is collected from $\{1000,250,70\}$ samples for system sizes $L=\{128,256,512\}$ respectively.}
    \label{fig:epm_cdf_collapse_final}
\end{figure*}

In this section we present results from the elastoplastic model (EPM) simulations. In Fig. \ref{fig:epm_cdf_collapse_final} we plot the distribution of failure times obtained from the EPM, where we have access to a very large number of samples due to a reduced computational cost and hence, a better estimate of the distribution function. We observe a satisfactory fit to the log-normal distribution irrespective of the annealing level (see Fig. \ref{fig:epm_cdf_collapse_final}(a) and Fig. \ref{fig:epm_cdf_collapse_final}(d)). Analogous to the atomistic results, we observe a satisfactory collapse of data across different $\gamma_{max}$ when the $x$-axis is scaled by the mean value (see Fig. \ref{fig:epm_cdf_collapse_final}(b) and Fig. \ref{fig:epm_cdf_collapse_final}(e)).

We observe that the scaled distribution depends on the degree of annealing, whereas a negligible dependence was observed in the atomistic case (for the large system size, Fig. \ref{fig:distribution_tf_anneal} (b)). This behaviour is observed for all the three system sizes (see Fig. \ref{fig:epm_cdf_collapse_final}(c) and Fig. \ref{fig:epm_cdf_collapse_final}(f)). For the well-annealed sample the ratio of standard deviation and mean ($\sigma/\mu$) of the log-normal distribution that describes the scaled data shows a decreasing trend with increasing system size, similar to what was observed in the atomistic case (compare with Fig. \ref{fig:distribution_tf_systemsize}(f)). For the poorly annealed sample this ratio ($\sigma/\mu$) does not change appreciably with system size, the value being nearly three to four times smaller than that for the well-annealed case. The reason for this difference between the EPM and atomistic simulation results, for the poorly annealed case, is not understood. 

\begin{figure*}[ht!]
    \centering
    \includegraphics[width=.85\linewidth]{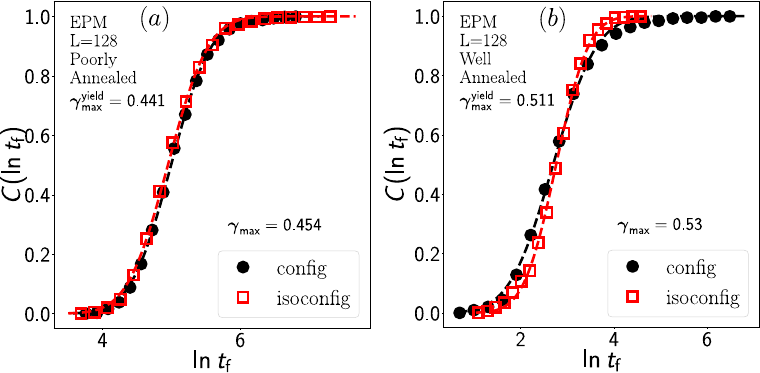}
    \caption{\textbf{Distribution of failure times for isoconfigurational runs in EPM:} Cumulative distribution of logarithm of failure times is plotted for runs with different initial conditions (in filled black circles) and for runs with the same initial condition with different random number seed (in red squares). In panel (a) we show the poorly annealed case and in panel (b) the well annealed case. System size is $L=128$, and $1000$ samples are considered.}
    \label{fig:epm_iso}
\end{figure*}

To check if stochasticity in the failure time arises from variations in the initial configurations, or from the dynamics, we perform isoconfigurational runs analogous to the atomistic case. We generate $1000$ trajectories from the same initial configuration by choosing a different random number generator seed for each trajectory. In Fig. \ref{fig:epm_iso}(a) we see that the isoconfigurational runs display the same standard deviation and mean of $\ln (t_f)$ when the sample is poorly annealed, while in Fig. \ref{fig:epm_iso}(b) where we show results for the well-annealed case, the standard deviation for the isoconfigurational runs is smaller by about $30 \%$. Thus we conclude that, similar to the atomistic case, a significant source of  stochasticity in the failure times is inherent in the dynamics, also in the case of the elastoplastic model. 

\subsection{Multiplicative damage accumulation, geometric Brownian motion, and failure} \label{Sec: multiplicativedamage}

\begin{figure*}[ht!]
        \centering
\includegraphics[width=.85\linewidth]{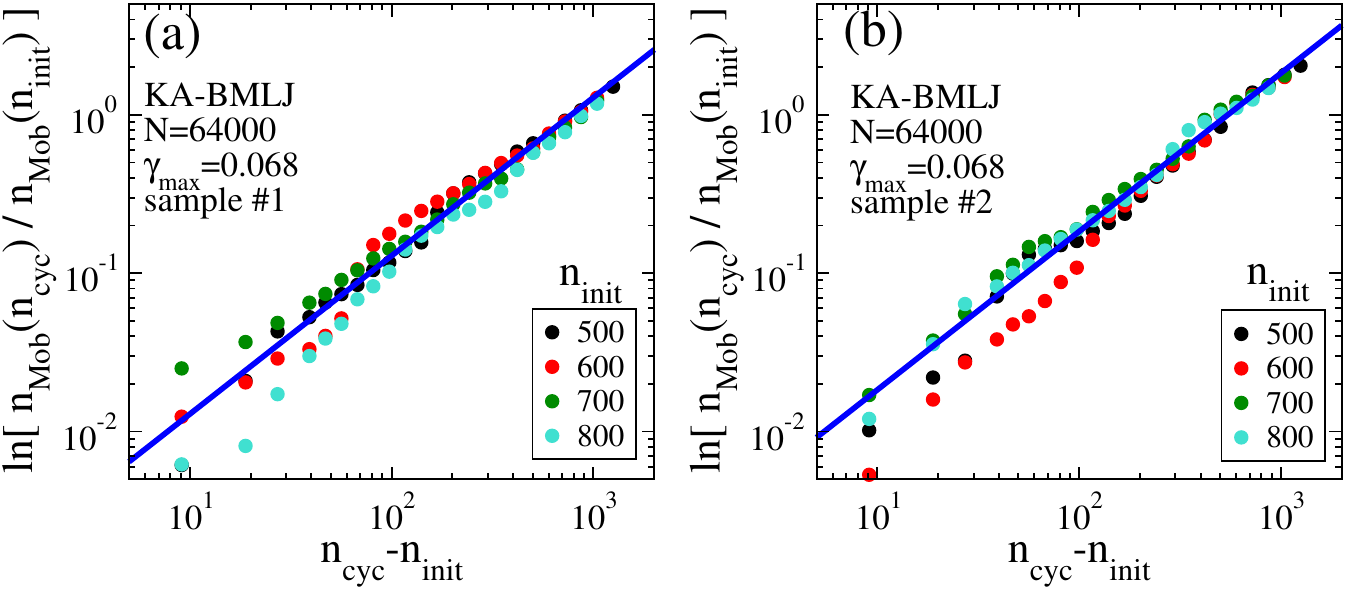}
    \caption{\textbf{Evolution of accumulated damage (KA-BMLJ):}  Evolution of accumulated damage with number of cycles in the atomistic simulation for two samples, for which the failure time $t_f=1850$ (a) and $t_f=1770$ (b). The straight line in each panel represents linear scaling. }
    \label{fig:multiplicative_damage}
\end{figure*}
\begin{figure*}[ht!]
        \centering  \includegraphics[width=.85\linewidth]{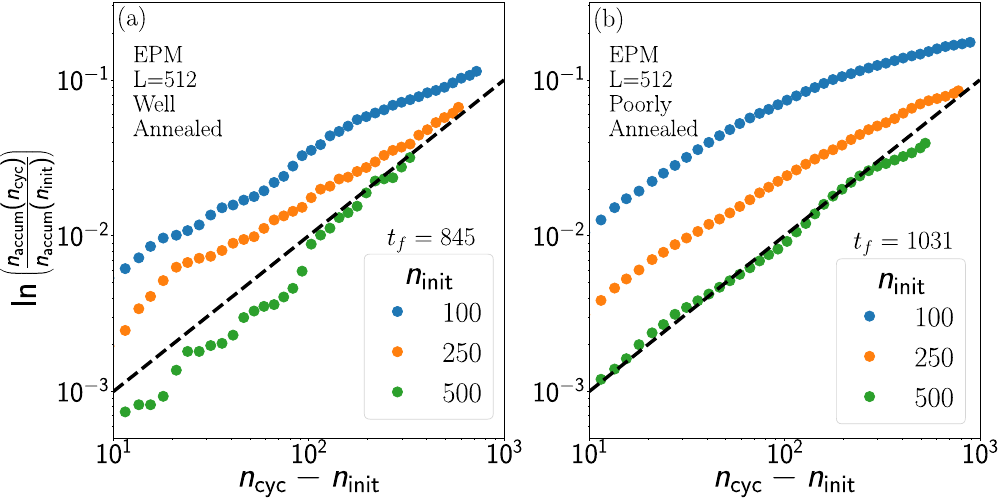}
    \caption{\textbf{Evolution of accumulated damage in EPM:} Evolution of fraction of sites that have undergone plastic activity for the (a) well-annealed sample and (b) poorly annealed sample. The dashed black line in each panel represents linear scaling. The failure time $t_f$ for the well-annealed sample is $t_f = 845$ cycles and for the poorly annealed sample it is $t_f = 1031$ cycles.}
\label{fig:epm_multiplicative_damage}
\end{figure*}

In previous work \cite{MaityNatPhys2026,khandare2026}, we found striking correlations between accumulated plasticity, in the form of the fraction of so-called ``mobile" particles, and dissipated work, and the failure times. In particular, the fraction of mobile particles was found in simulations \cite{MaityNatPhys2026} to reach a threshold value at the time of failure, largely independently of the strain amplitude and sample-to-sample variations (similar results, but with differences, were observed in the EPM \cite{khandare2026}). This suggests the picture of an accumulation of damage to a threshold leading to failure, but does not directly offer insight into the process involved. In the following, we consider a preliminary analysis of our results, within the framework of multiplicative degradation of materials, commonly used to describe the lifetime of materials before failure \cite{ParkIEEE05,ParkLDA05,Taloni2018} 
With a stochastic model of dynamics for some quantification of damage, which is multiplicative, the hitting time to reach the threshold damage can be taken as the failure time of the material. In general, one can write cumulative damage at time $n+1$ as,
\begin{align}
    x_{n+1}=x_n+\xi_n h(x_n),
\end{align}
where $\xi_n$ is the independent and identically distributed random variable and $h(n)$ is called damage model function. For $h(x_n)=1$ the model is additive, and for $h(x_n)=x_n$, the model is multiplicative and is called the geometric Brownian process \cite{ParkLDA05,ParkIEEE05} 
when $\xi_i$ has a positive drift (which will depend on the properties of $\xi_n$). The accumulated damage can be written as 
\begin{align}
    x_{n}=x_0 \prod_{i=0}^n(1+\xi_i)
\end{align}
or,
\begin{align}
    \ln\left(\frac{x_n}{x_0}\right)=\sum_{i=1}^n\ln(1+\xi_i)
\end{align}
where $x_0$ is the initial damage of the process. Using the central limit theorem, one can get the distribution of $\ln\left( \frac{x_n}{x_0} \right)$ as,
\begin{align}
    P \left(\ln\left( \frac{x_n}{x_0} \right )\right ) = \mathcal{N}(n\mu_\eta, n\sigma^2_\eta)
\end{align}
where $\mu_\eta$ and $\sigma_\eta^2$ are the mean and standard deviation of the variable $\eta=\ln(1+\xi_i)$, respectively. Hence, the mean of $\ln \left( \frac{x_n}{x_0} \right)$ will grow linearly with $n$. From the first passage time problem, one can show that the distribution of failure times (time to reach a constant threshold value of accumulated damage at which the system fails) follows the inverse Gaussian distribution \cite{ParkLDA05}.

We define the accumulated damage in atomistic simulations to be $n_{Mob}(n_{cyc})$, the  accumulated fraction of mobile particles -- that move non-affinely a large distance ($D^2_{min}>1.2$) \cite{MaityNatPhys2026} -- up to $n_{cyc}$ cycles of shear. We show the growth of accumulated damage $\ln [ n_{Mob}(n_{cyc})/n_{Mob}(n_{init})]$
with the shear cycle in Fig. \ref{fig:multiplicative_damage}(a) and \ref{fig:multiplicative_damage}(b) for a different choice of $n_{init}$. Data clearly indicates linear growth, which allows us to infer that the damage accumulates in fatigue failure in glasses in a multiplicative way, which can be modelled through a stochastic degradation process. Analogous results from the EPM are shown in Fig. \ref{fig:epm_multiplicative_damage}(a) and Fig. \ref{fig:epm_multiplicative_damage}(b) where $n_{accum}(n_{cyc})$ is defined as the accumulated fraction of sites that have undergone a plastic event till cycle number $n_{cyc}$. Linear growth is particularly apparent as $n_{init}$ is increased, indicating that the damage mechanism (barring early cycles) in our EPM is multiplicative in nature.



\section{Conclusions}
\label{Sec: conclusions}

In summary, we have investigated the stochastic nature of fatigue failure times in glasses subjected to cyclic shear, using both atomistic simulations and an elasto-plastic model (EPM). The failure times are observed to be well described by the lognormal distribution. Interestingly, the cumulative distributions of the logarithm of failure times at different strain amplitudes, $\gamma_{max}$, collapse onto a single master curve when the horizontal axis is scaled by the mean value, indicating that the standard deviation is proportional to the mean value. This observation holds for different model glasses — the KA-BMLJ and CP models — as well as for the EPM we study. The data collapse across $\gamma_{max}$ applies also for different system sizes, although the ratio of the standard deviation to the mean of the logarithm of failure times decreases with increasing system size. The exact form of this decrease remains unclear, and further analysis across more system sizes is required to understand the behaviour in the thermodynamic limit. A similar data collapse is observed across different degrees of annealing in atomistic simulations; however, the EPM results deviate, exhibiting different ratios depending on the degree of annealing. Finally, we compare distributions of failure times from iso-configurational runs (same initial configuration, different velocities).
The resulting distributions are identical to those obtained from ensembles of initial configurations, indicating that the stochasticity of failure times is inherent to the dynamics of fatigue failure rather than merely arising from the distribution of initial configurations. 

To understand the origin of these lifetime statistics, we analyze the evolution of accumulated damage and find that its logarithm grows approximately linearly with cycle number both for the model glasses and the elastoplastic model, indicating that fatigue failure in these systems can be interpreted as a multiplicative damage-accumulation process. Within a stochastic degradation framework, taking the incremental damage as a stochastic variable with positive drift provides the distribution of first passage times to reach a threshold damage, which is an inverse-Gaussian–type distribution, which also captures the observed failure time statistics well.

Together, these results describe the stochastic nature of the fatigue failure phenomena, their dependence on various system properties, and the dynamic origin. Preliminary analysis supports the description of damage accumulation to be a multiplicative process, but without specific reference to microscopic processes and causes. Further investigations along these lines are required to gain further insights into a microscopic picture, which may provide a better understanding of the fatigue failure process and useful guidance for fatigue life predictions in experimental systems.

\section*{Acknowledgments} We thank R. L. Jack and Debargha Sarkar for useful discussions. We acknowledge the National Supercomputing Mission facility (Param Yukti) at the Jawaharlal Nehru Centre for Advanced Scientific Research for computational resources. S.S. acknowledges support from the Science and Engineering Research Board (Anusandhan National Research Foundation) (India) through the JC Bose Fellowship (Grant No. JBR/2020/000015) and a grant under Scientific and Useful Profound Research Advancement (SUPRA) (Grant No. SPR/2021/000382).



\providecommand{\newblock}{}

\clearpage
\newpage

\end{document}